%% file: Hamiltonian_Yang-Mills.tex
\documentclass[11pt,a4paper]{article}
\usepackage{amsmath,amsfonts,amssymb,amscd,graphicx}
\catcode`\@=11
\@addtoreset{equation}{section}

\catcode`\@=12

\newtheorem{Definition}{Definition}[section]
\newtheorem{Proposition}{Proposition}[section]

\newfont{\gotico}{eufm10 scaled\magstephalf}
\newfont{\qvd}{msam10 scaled\magstephalf}

\def\frak#1{\mbox{\gotico #1}}

\def\demo{\par\noindent{\sc Proof. }\begingroup}
\def\enddemo{\hskip1em \mbox{\qvd \char3}\endgroup\par\medskip}

\def\iff{\Leftrightarrow}
\def\interior{\,\hbox{\vrule depth0pt height.6pt width4pt%
\vrule depth0pt height8pt}\;\,}

\def\de#1/de#2{\frac{\partial {#1}}{\partial {#2}}}
\def\De#1/de#2{\dfrac{\partial {#1}}{\partial {#2}}}

\def\det{{\rm det}\,}

\def\L{{\cal L}}
\def\H{{\cal H}}

\def\je{{\cal J}\/(E)}

\def\he{\H\/(E)}
\def\pe{\Pi\/(E)}

\begin{document}
\vskip-2cm

\title{On the Hamiltonian formulation of Yang--Mills gauge theories}

\author{Stefano Vignolo, Roberto Cianci\\
        DIPTEM Sez. Metodi e Modelli Matematici, Universit\`a di Genova \\
        Piazzale Kennedy, Pad. D - 16129 Genova (Italia)\\ 
        E-mail: vignolo@diptem.unige.it, cianci@diptem.unige.it
\and  
       Danilo Bruno\\
       Dipartimento di Matematica, Universit\`a di Genova \\
                Via Dodecaneso 35 - 16146 Genova (Italia) \\
                E-mail: bruno@dima.unige.it
}

\date{}              
\maketitle

\begin{abstract}{The Hamiltonian formulation of the theory proposed in \cite{CVB1,CVB2}
is given both in the Hamilton--De Donder and in the Multimomentum Hamiltonian geometrical approaches.
$(3+3)\/$ Yang--Mills gauge theories are dealt with explicitly in order to restate them
in terms of Einstein--Cartan like field theories. 
}
\end{abstract}

\section{Introduction}
\input{par_0}
\section{The geometrical framework}
\input{par_1}
\section{The Hamiltonian framework}
\input{par_2}
\section{Multimomentum Hamiltonian formulation}
\input{par_3}
\section{$3+3$ Yang--Mills fields}
\input{par_4}

\input{ref}
\end{document}

%% file: par_0.tex
\noindent In some of our recent works \cite{CVB1,CVB2,VC} a new geometrical framework
for Yang--Mills field theories and General Relativity in the tetrad--affine formulation
has been developed.

\noindent The construction of the new geometrical setting started from the observation
that even though the Lagrangian densities of the above theories are defined over the
first jet--bundle of the configuration space, they only depend of the antisymmetric
combination of field derivatives in the space--time indexes. As for the Yang--Mills case,
this is the reason for the singularity in the Lagrangian.

\noindent The idea consists is considering a suitable quotient of the first jet--bundle,
making two sections equivalent when they possess a first order contact with respect to
the exterior covariant differentiation, instead of the whole set of derivatives. The
fiber coordinates of the resulting quotient bundle are the antisymmetric combinations of
the field derivatives that appear in the Lagrangian.

\noindent The geometry of the new space has been widely studied, in order to build as
many usual geometric structures of the jet--bundle theory as possible, such as contact
forms, jet--prolongations of sections, morphisms and vector fields. These are the
geometric tools that are needed to implement variational problems in the
Poincar\'e--Cartan formalism. Moreover, particular choices for
the fiber coordinates have been shown to be possible: they consist in the components of
the strength tensor for Yang--Mills theories and in the torsion and curvature tensors for
General Relativity.

\noindent This resulted into the elimination of some un--physical degrees of freedom from
the theory (represented by un-necessary jet--coordinates) and to even obtain a {\it
regular Lagrangian} theory in the case of Yang--Mills fields.

\noindent This last consideration was the thrust that moved us to write the present work:
the presence of a regular Lagrangian allows us to write a Hamiltonian version of the
theory proposed in \cite{CVB1,CVB2}. The advantages arising from the present approach,
with respect to the already existing formulations, based on singular Lagrangians (compare, for example,
with \cite{Sardanashvily,Mangiarotti}), are striking. In fact, the singularity of the
Lagrangian is the source of known drawbacks: the equations are defined on a constraint
sub--manifold, there exist multiple Hamitonian forms associated with the same Lagrangian
and the equivalence between Euler--Lagrange and Hamilton equations is not a direct
consequence of Legendre transform any more.

\noindent On the contrary, the situation in the new geometrical framework is simpler and
more elegant: the ``Lagrangian'' space and the phase--space have the same dimension and
the Legendre transform is a (local) diffeomorphism. This ensures the direct equivalence
of Lagrangian and Hamiltonian formulations, both in the Hamilton--De Donder (section 3)
and in the Multimomentum Hamiltonian approach (section 4).

\noindent Finally, we devoted the last section to study the peculiar $(3+3)$ Yang--Mills
theory. Starting from the work made in \cite{Raiteri1,Raiteri2}, we showed that a
coordinate transformation in the phase--space, together with the Poincar\'e--Cartan
approach in our new formalism, allows to describe a (3+3) Yang--Mills theory by means of
Einstein--Cartan like equations in 3 dimensions. In particular, in the case of a free $(3+3)$
Yang--Mills field the geometrical construction gives rise to a sort of first--order purely
frame--formulation of a General Relativity like theory.

\noindent
This result is interesting for its
further developments: in fact we will show in a subsequent work that an analogous geometrical machinery
may be applied to build a {\it first--order} 
purely frame--formulation of General Relativity in four dimensions \cite{VCB1}.

%% file: par_1.tex
The present section is devoted to revising the geometrical structure that has 
been introduced in \cite{CVB1,CVB2} to describe Yang--Mills theories. 

Let $\pi : P \to M$ be a principal fiber bundle, with structural group $G$ and
let $x^i,g^\mu$ denote a system of local fibered coordinates on $P$.
$J_1\/(P)$ denotes the first jet-bundle of $\pi : P \to M$ and it is referred to
local coordinates $x^i,g^\mu,g^\mu_i\left(\simeq \de g^\mu /de{ x^i} \right)$.

The space of principal connections on $P$ is identified with the quotient bundle
$E := J_1(P)/G$ with respect to the (jet-prolongation of) the right action
$R_h$ of the structural group on $P$. If $V^\mu_\nu(g,h)$ 
represents the differential of the right multiplication $R_h$ in $g\in G$, a set 
of local coordinates in the quotient space is provided by $x^\mu,a^\mu_i = - 
g_i^\nu V_\nu^\mu(g,g^{-1})$, subject to the following transformation laws:
\begin{equation}\label{1.1}
\bar{x}^i =\bar{x}^i\/(x^j), \qquad \bar{a}^\mu_i = \left[ 
Ad\/(\gamma^{-1})^\mu_\nu a^\nu_j + 
W^\mu_\nu\/(\gamma^{-1},\gamma)\de\gamma^\nu/de{x^j} \right]\de 
x^j/de{\bar{x}^i}
\end{equation}
where $Ad^\mu_\nu\/$ and $W^\mu_\nu\/$ denote respectively the adjoint representation 
of $G\/$ and the differential of the left multiplication in $G\/$, while
$\gamma :U\subset M\to G\/$ ($U\/$ open set) is an arbitrary smooth map.

As a consequence, the bundle $E\to M$ has the nature of an affine 
bundle, whose sections represent principal connections over $P\to M$. In fact, 
every section $\omega:M\to J_1(P)/G$ yields a connection $1$-form on $P$, 
locally described as:

\begin{equation}\label{1.2}
\omega\/(x,g) = \omega^\mu\/(x,g) \otimes \underbar{e}_\mu := \left[
Ad\/(g^{-1})^\mu_\nu a^\nu_i\/(x)\,dx^i + W^\mu_\nu\/(g^{-1},g)\;dg^\nu
\right]\otimes \underbar{e}_\mu
\end{equation}
where $\underbar{e}_\mu\/$ ($\mu =1,\ldots,r\/$) indicate a basis of the Lie algebra $\frak g\/$ of $G\/$.

Finally, let $\hat \pi : J_1\/(E)\to E$ be the first jet--bundle
associated with the bundle $E\to M$, described by the set of local coordinates
$x^i,a^\mu_i,a^\mu_{ij}\left(\simeq \de a^\mu_i /de {x^j} \right)$. 

In order to
provide a better geometrical framework to describe Yang-Mills gauge theories,
the following equivalence relation is introduced in $J_1\/(E)$: let
$\omega_1 = (x^\mu,a^\mu_i,a^\mu_{ij}),\omega_2 = (x^\mu,a^\mu_i,\hat
a^\mu_{ij}) \in J_1(E)$ be such that $\hat \pi (\omega_1) = \hat \pi
(\omega_2)$, then:
\begin{equation}\label{1.2tris}
\omega_1 \sim \omega_2 \quad \iff \quad (a^\mu_{ij}-a^\mu_{ji}) =
(\hat a^\mu_{ij}-\hat a^\mu_{ji})
\end{equation}
This means that two sections are declared equivalent if their skew--symmetric
derivatives are equal. In more geometric terms, being every section of the
bundle $E\to M$ represented by a connection 1--form, the first jet--bundle has 
been constructed assuming that the equivalence between sections having a 
first--order contact is evaluated through the exterior differentiation (or, equivalently, 
the covariant exterior differentiation), instead 
of the whole set of partial derivatives. 

Let $\je := J_1\/(E)/\sim$ denote the quotient bundle with respect to the 
above defined equivalence relation and $\rho : J_1\/(E) \to \je$ the 
canonical (quotient) projection. The bundle $\je$ is endowed with a set of local 
fibered coordinates $x^i,a^\mu_i,A^\mu_{ij}:= \frac{1}{2} 
\left(a^\mu_{ij} -a^\mu_{ji}\right) (i<j)$, subject to the following 
transformation laws:
\begin{equation}\label{1.2bis}
\bar{A}^\mu_{ik} = \de x^j/de{\bar{x}^i}\de x^h/de{\bar{x}^k} \left[ 
Ad\/(\gamma^{-1})^\mu_\nu A^\nu_{jh} + \frac{1}{2}\/\left( 
\de{Ad\/(\gamma^{-1})^\mu_\nu}/de{x^h} a^\nu_j - 
\de{Ad\/(\gamma^{-1})^\mu_\nu}/de{x^j} a^\nu_h \right) +\frac{1}{2}\/\left( 
\de\eta^\mu_j/de{x^h} - \de\eta^\mu_h/de{x^j} \right)\right]
\end{equation}
where 
$\eta^\mu_j\/(x):=W^\mu_\nu\/(\gamma^{-1}\/(x),\gamma\/(x))\de\gamma^\nu\/(x)/de
{x^j}\/$.

This newly defined geometrical framework is endowed with the most 
common features provided by a standard jet--bundle structure.

\bigskip
\noindent $\bullet\/$ {\it $\cal{J}$-extension of sections}. Given a section $\sigma : 
M\to E$ its ${\cal J}$-extension is defined as ${\cal J} \sigma := \rho \circ 
J_1 \sigma :M \to \je$, namely projecting the standard jet--prolongation to 
$\je$ by means of the quotient map. Conversely, every section $s:M\to \je$ will 
be said to be holonomic if there  exists a section $\sigma:M\to E$ such that $s 
= {\cal J}\sigma$.

\bigskip
\noindent $\bullet\/$ {\it Contact forms}. Let us define the following 
$2$-forms on $\je$:
\begin{equation}\label{1.3}
\theta^\mu := da^\mu_i \wedge dx^i + A^\mu_{ij} dx^i\wedge dx^j
\end{equation}
They undergo the transformation laws $\bar{\theta}^\mu = Ad\/(\gamma^{-1})^\mu_\nu \theta^\nu\/$.
The vector bundle which is locally spanned by the $2$-forms \eqref{1.3} will be 
called the contact bundle ${\cal C}(\je)$ and any section $\eta:\je\to {\cal 
C}(\je)$ will be called a contact $2$-form. Contact forms are such that 
$s^*(\eta)=0$ whenever $s :M\to \je$ is holonomic. 

\bigskip
\noindent $\bullet\/$ {\it $\cal{J}$-prolongations of morphisms and vector fields}. A 
generic morphism $\Phi:E\to E$, fibered over $M$, can be raised to a morphism 
${\cal J} \Phi : \je \to \je$ considering its ordinary jet--prolongation and 
restricting it to $\je$ through the quotient map, namely:
\[
{\cal J} \Phi (z) := \rho \circ j_1\Phi\/(w) \quad \forall\; w \in 
\rho^{-1}\/(z)\;,\; z\in \je
\]
As a matter of fact, not every morphism $\Phi:E\to E$ commutes with the quotient 
map and produces a well defined prolongation (i.e. independent of the choice of 
the representative in the equivalence class), but it has to satisfy the 
condition:
\begin{equation}\label{1.4}
\rho\circ j_{1}\Phi\/(w_1)=\rho\circ j_{1}\Phi\/(w_2) \qquad\forall\;w_1,w_2 \in 
\rho^{-1}\/(z)
\end{equation}
Referring to \cite{CVB1} for the proof, it is easy to see that the only 
morphisms satisfying condition \eqref{1.3} are necessarily of the form:
\begin{equation}\label{1.5}
\left\{
\begin{array}{l}
y^i = \chi^i\/(x^j)\\
\\
b^\nu_i = \Phi^\nu_i\/(x^j,a^\mu_j)= \Gamma^\nu_\mu\/(x) \De x^r/de{y^i} a^\mu_r 
+ f^\nu_i\/(x)
\end{array} 
\right.
\end{equation}
where $\Gamma^\nu_\mu\/(x)\/ $ and $f^\nu_i\/(x)\/$ are arbitrary local functions 
on $M\/$. Their ${\cal J}-$prolongation is:
\[
\left\{
\begin{array}{l}
y^i = \chi^i\/(x^k)\\
\\
b^\nu_i = \Gamma^\nu_\mu\/(x) \De x^r/de{y^i} a^\mu_r + f^\nu_i\/(x)\\
\\
B^\nu_{ij} = \Gamma^\nu_{\mu}A^\mu_{ks}\De x^k/de{y^i}\De x^s/de{y^j} + 
\frac{1}{2}\left[\De\Gamma^\nu_\mu/de{x^k}\left(\De x^k/de{y^j}\De x^r/de{y^i} - 
\De x^k/de{y^i}\De x^r/de{y^j}\right)a^\mu_r + \De f^\nu_i/de{x^k}\De 
x^k/de{y^j} - \De f^\nu_j/de{x^k}\De x^k/de{y^i}\right]\end{array} \right.
\]
In a similar way (compare with \cite{CVB1}), it is easy to prove that the only 
vector fields of the form
\begin{equation}\label{1.6} 
X=\epsilon^i\/(x^j)\,\de /de{x^i} + \left(-\de{\epsilon^k}/de{x^q}a^\mu_k + 
D^\mu_\nu\/(x^j)a^\nu_q + G^\mu_q\/(x^j)\right)\,\de /de{a^\mu_q}
\end{equation}
(where $\epsilon^i\/(x)\/$, $D^\mu_\nu\/(x)\/$ and $G^\mu_q\/(x)\/$ are arbitrary local functions on $M\/$) 
can be ${\cal J}-$prolonged to vector fields over $\je$ as follows:
\begin{equation}\label{1.7}
{\cal J}\/(X)\/(z):=\rho_{* \rho^{-1}\/(z)}(j_1\/(X)) \qquad\forall\, z\in {\cal 
J}\/(P)
\end{equation}
The resulting vector field has the form:
\[
{\cal J}\/(X)=\epsilon^i\/(x^j)\,\de /de{x^i} + 
\left(-\de{\epsilon^k}/de{x^q}a^\mu_k + D^\mu_\nu\/(x^j)a^\nu_q + 
G^\mu_q\/(x^j)\right)\,\de /de{a^\mu_q} + \sum_{i<j}h^\mu_{ij}\,\de 
/de{A^\mu_{ij}}
\]
where
\[
h^\mu_{ij}= \frac{1}{2}\left( \de D^\mu_\nu/de{x^j}a^\nu_i - \de 
D^\mu_\nu/de{x^i}a^\nu_j + \de G^\mu_i/de{x^j} - \de G^\mu_j/de{x^i}\right) + 
D^\mu_{\nu}A^\nu_{ij} + \left( A^\mu_{ki}\de{\epsilon^k}/de{x^j} - 
A^\mu_{kj}\de{\epsilon^k}/de{x^i}\right)
\]

\noindent Finally, in order to adapt the geometrical framework to the presence 
of the covariant differentiation induced by connections, it is useful to 
introduce a set of new fibered local coordinates over $\je$ of the form:
\begin{equation}\label{1.8}
x^i = x^i \quad a^\mu_i = a^\mu_i \quad F^\mu_{ij} = 2A^\mu_{ji} + a^\nu_j 
a^\rho_i C^\mu_{\rho\nu}
\end{equation}
where $C^\mu_{\rho\nu}$ are the structure coefficients of the group $G$. The 
latter are subject to the following transformations laws:
\begin{equation}
\label{1.9}
\bar{F}^\mu_{ik} = \de x^j/de{\bar{x}^i}\de x^h/de{\bar{x}^k} 
Ad\/(\gamma^{-1})^\mu_\nu F^\nu_{jh} 
\end{equation} 

\noindent Using the new coordinates, every Yang--Mills Lagrangian $m$--form can 
be expressed as 
\begin{equation}
\label{1.10}
L = \L\/(x^i,a^\mu_i,F^\mu_{ij})\; ds 
\end{equation} 
Moreover, it is possible to define a corresponding Poincar\'e--Cartan $m$-form 
over $\je$, expressed as
\begin{equation}
\label{1.11}
\Theta_L := \L \; ds - \frac{1}{2} \theta^\mu \wedge P_\mu 
\end{equation} 
where $P_\mu := \de{\cal L}/de{F^\mu_{ij}}\,ds_{ij}\/$, $ds_{ij}:= \de 
/de{x^i}\interior \de /de{x^j}\interior ds\/$.

\noindent The presence of the Poincar\`e--Cartan form allows to deduce the evolutions
equations for Yang--Mills fields looking for the stationary points of the functional
\begin{equation}\label{1.12}
    A_L\/(\gamma) := \int_D \gamma^*\/(\Theta_L) \quad \forall\;\gamma:D\subset M \to \je
\end{equation}

\noindent The stationarity condition for $A_L$ (taking null variations at the boundary of
the compact domain $D$) is equivalent to the conditions (compare with \cite{CVB1,CVB2}):
\begin{subequations}\label{1.13}
\begin{equation}
    \gamma^*\/(\theta^\mu) = 0
\end{equation}
\begin{equation}
    \gamma^*\/\left(\de \L /de {a^\mu_i} - D_j \de \L /de{F^\mu_{ji}}\right) = 0
\end{equation}
\end{subequations}

\noindent The first equation ensures the kinematic admissibility of the critical section
$\gamma$, while the second represents the field equations of the problem. As a matter of
fact, the kinematical admissibility is directly obtained from the variational principle
and is not imposed as an a-priori condition, as a consequence of the {\it regularity} of
the Lagrangian $\L$ within the new framework provided by $\je$. 

%% file: par_2.tex
Let $\Lambda^m\/(E)$ denote the modulus of $m$-forms over $E$, and let
$\Lambda_r^m\/(E)\subset \Lambda^m\/(E) $ ($r<m$) be the sub-bundle consisting
of those $m$-forms on $E$ vanishing when $r$ of its given arguments are vertical
vectors over the bundle $E\to M$. It is obvious that the above defined bundles
form a chain of vector bundles over $E$ such that:
\[
0 \subset \Lambda_1^m(E)\subset \Lambda_2^m(E) \subset \ldots \subset
\Lambda_r^m(E) \subset \ldots \subset \Lambda^m(E)
\]
In particular the attention will be focussed on the first two sub-spaces. Given
a system of local coordinates over $E$ and let $ds = d\/x^1\wedge\ldots\wedge
d\/x^m$, they can be respectively described as:
\begin{equation}
\label{2.1}
\Lambda_1^m(E):=\{\omega\in\Lambda^m(E) : \omega = p\,ds\}
\end{equation}
and
\begin{equation}
\label{2.2}
\Lambda_2^m(E):=\{\omega\in\Lambda^m(E) : \omega = p\,ds + \Pi_\mu^{ji}\,d\/a^\mu_{\,i}\wedge d\/s_j\}
\end{equation}
where $d\/s_j := \de /de {x^j} \interior ds$. It is then possible to assume
$\{x^i,a^\mu_i,p\}$ as a system of local coordinates on $\Lambda_1^m\/(E)$, subject to
the transformations laws $p= J \bar p$ (where $J:= \det\left\|\de \bar x^i /de {
x^k}\right\|$).

\noindent A set of local coordinates for  $\Lambda_2^m\/(E)$ is provided by the functions
$\{x^i,a^\mu_i,p,\Pi_\mu^{ij} \}$. The latter are subject to a set of transformation laws
described by eqs.~\eqref{1.1}, together with:
\begin{subequations}
\label{2.3}
\begin{equation}
p = J\/\left( \bar p + \bar \Pi_\mu^{ji} \left(\de Ad(\gamma^{-1})^\mu_\nu /de {x^q}
a^\nu_p + \de\eta^\mu_p/de{x^q} \right) \de x^q/de{\bar{x}^j}\de x^p/de{\bar{x}^i} + \bar
\Pi^{ij}_\mu (Ad\/(\gamma^{-1})^\mu_\nu a^\nu_p + \eta^\mu_p) \frac{\partial^2
x^p}{\partial \bar x^j \partial \bar x^i}\right)
\end{equation}

\begin{equation}
\Pi_\nu^{pq} = \bar \Pi_\mu^{ij} Ad(\gamma^{-1})^\mu_\nu \de x^q/de{\bar{x}^j}\de
x^p/de{\bar{x}^i}J
\end{equation}
\end{subequations}

\noindent The bundle $\Lambda_2^m(E)$ is endowed with the canonical Liouville $m$-form,
locally expressed as
\begin{equation}
\label{2.4} \Theta := p\; ds + \Pi_\mu^{ji} \;d\/a^\mu_{\,i}\wedge d\/s_j
\end{equation}
whose differential
\begin{equation}
\label{2.5} \Omega := d\Theta = dp \wedge ds + d \Pi_\mu^{ji}\wedge d\/a^\mu_{\,i}\wedge
d\/s_j
\end{equation}
is a multisymplectic $(m+1)-$form over $\Lambda_2^m\/(E)$.

\noindent A deeper geometrical insight in the problem can be given observing that
eqs.~\eqref{2.3} make $\Lambda_1^m$ into a vector sub-bundle of $\Lambda_2^m$, thus
allowing  us to introduce the quotient bundle $\Lambda_2^m/\Lambda_1^m$. As a consequence
of the definition, the latter has the nature of a vector bundle over $E$ and is locally
described by the system of coordinates $x^i,a^\mu_i,\Pi_\mu^{ij}$. It is worth noticing
that the transformation law (\ref{2.3}a) makes $\pi: \Lambda_2^m\/(E)\to
\Lambda_2^m/\Lambda_1^m$ into an affine bundle.

\noindent The {\it phase space} is defined as the vector sub--bundle $\pe \subset
\Lambda_2^m\/(E)/\Lambda_1^m\/(E)$ consisting of those elements
$z\in\Lambda_2^m\/(E)/\Lambda_1^m\/(E)$ satisfying the requirement
\begin{equation}\label{2.4bis}
    \Pi_\mu^{ij} (z) = - \Pi_\mu^{ji} (z)
\end{equation}

\noindent Condition \eqref{2.4bis} is well--posed because of the transformation laws
\eqref{2.3}. A local system of coordinates for $\pe$ is provided by
$x^i,a^\mu_i,\Pi_\mu^{ij}(i<j)$, subject to the same transformation laws (\ref{2.3}b).
Besides, being $\pe$ a vector sub--bundle, the immersion $i :\pe \to
\Lambda_2^m\/(E)/\Lambda_1^m\/(E)$ is well defined and is locally represented by
eq.~\eqref{2.4bis} itself.

\noindent The pull--back bundle $\hat \pi:\he\to\pe$ defined by the following commutative
diagram

\begin{equation}\label{2.4tris}
\begin{CD}
\he     @>{\hat i}>>    \Lambda_2^m\/(E)    \\
@V{\hat \pi}VV              @VV{\pi}V  \\
\pe        @>i>>     \Lambda_2^m\/(E)/\Lambda_1^m\/(E)
\end{CD}
\end{equation}
will now be taken into account. A local coordinate system for $\he$ is provided by
$x^i,a^\mu_i,\Pi_\mu^{ij}(i<j),p$, subject to transformation laws (\ref{2.3}b), together
with:
\begin{equation}\label{2.4i}
    p = J \/ \left( \bar p + \bar \Pi_\mu^{ji} \left(\de Ad(\gamma^{-1})^\mu_\nu /de {x^q}
a^\nu_p + \de\eta^\mu_p/de{x^q} \right) \de x^q/de{\bar{x}^j}\de x^p/de{\bar{x}^i}
\right)
\end{equation}
the latter being the antisymmetric part of eq.~(\ref{2.3}a). The above transformation law
shows that the bundle $\hat \pi : \he\to\pe$ has the nature of an affine bundle over the
phase space. Every section $h:\pe \to \he$ will be called a {\it
Hamiltonian section}, and will be locally described in the form:
\begin{equation}
\label{2.6} h: p = -\H\/(x^i,a^\mu_i,\Pi_\mu^{ij})
\end{equation}

\noindent The presence of the immersion $\hat i:\he\to\Lambda_2^m\/(E)$, endows $\he$
with the canonical $m$-form $\hat i^*\/(\Theta)$, locally expressed as in
eq.~\eqref{2.4}. The latter will be simply denoted as $\Theta$ and will be called the
Liouville form on $\he$.

\noindent The presence of the $m-$form $\Theta$ on $\he$, allows to create a
correspondence between the Hamiltonian and the Lagrangian viewpoints, based on the
existence of a unique diffeomorphism $\lambda : \je \to \he$ fibered over $E$ satisfying
the requirement:
\begin{equation}
\label{2.8}
\Theta_L = \lambda^*\/(\Theta)
\end{equation}
Such a diffeomorphism will be called the {\it Legendre map}. Given a set of local
coordinates $x^i,a^\mu_i,F^\mu_{ij} (i<j)$ on $\je$ and $x^i,a^\mu_i,\Pi_\mu^{ij}(i<j),p$
on $\he$, and taking eqs.~\eqref{1.8} and \eqref{1.11} into account, the
Poincar\'e--Cartan m-form can written as
\begin{equation} \label{2.9}
\Theta_L = \left(L -
\frac{1}{2}\left(F^\mu_{kr} + a^\nu_k a^\rho_r C^\mu_{\rho\nu}\right)\de L/de
{F^\mu_{kr}} \right) ds +  \de L/de {F^\mu_{rk}} d a^\mu_k \wedge ds_r
\end{equation}
and the Legendre map defined by eq.~\eqref{2.8} is such that:
\begin{equation}
\label{2.10}
\lambda: \/ \left\{
\begin{array}{l}
x^i = x^i \\
\\
a^\mu_i = a^\mu_i \\
\\
p\/(x^j,a^\alpha_j,F^\alpha_{ij}) = L - \frac{1}{2}\left(F^\mu_{kr}+a^\nu_k
a^\rho_r C^\mu_{\rho\nu}\right)\de L/de {F^\mu_{kr}}  \\
\\
\Pi_\mu^{ij}\/(x^j,a^\alpha_j,F^\alpha_{ij}) = \de L/de {F^\mu_{ij}}
\end{array}
\right.
\end{equation}

\noindent The most striking feature of the Legendre transformation between $\je$ and
$\he$ is provided by its regularity, due to the acquired regularity of the Yang--Mills
Lagrangian in the space $\je$. In particular the condition
\[
\det\left(\de \Pi_\mu^{ij} /de {F^\alpha_{rk}}\right) \neq 0 \quad \forall \;
i<j \; , \; r<k \quad \forall\; \alpha ,\mu
\]
assures the local invertibility of the last equation \eqref{2.10}, allowing to obtain the
coordinates $F^\alpha_{ij}$ as functions $F^\alpha_{ij} =
F^\alpha_{ij}\/(x^j,a^\alpha_j,\Pi_\mu^{ij})$. Thus, the Legendre map has the nature of a
regular immersion of $\je$ into $\he$, yielding a submanifold $\lambda\/(\je)\subset
\he$, locally described by:
\begin{equation}
\label{2.10bis}
p\/(x^j,a^\alpha_j,\Pi_\mu^{ij}) = L\/(x^j,a^\alpha_j,\Pi_\mu^{ij}) -
\frac{1}{2}\left(F^\mu_{kr}\/(x^j,a^\alpha_j,\Pi_\mu^{ij})+a^\nu_k a^\rho_r
C^\mu_{\rho\nu}\right)\Pi_\mu^{kr}
\end{equation}

\noindent In accordance with the literature, the function
\begin{equation}
\label{2.10tris} H(x^i,a^\mu_i,\Pi^\mu_{ij})= - L (x^i,a^\mu_i,\Pi^\mu_{ij}) +
\frac{1}{2}F^\mu_{kr}(x^i,a^\mu_i,\Pi_\mu^{ij})\Pi_\mu^{kr}
\end{equation}
will be called the {\it Hamiltonian} of the system.

\noindent If the phase space $\pe$ is taken into account, the composition $\hat \lambda
:= \hat \pi \circ \lambda : \je \to \pe$ results to be a (local) diffeomorphism. As a
consequence, its (local) inverse map $\hat\lambda^{-1}:\pe \to \je$ can be considered. Taking the
derivatives of eq.~\eqref{2.10tris} with respect to $\Pi_\mu^{ij}$ and using the
antisymmetric properties of the coordinates, one gets the coordinate representation for
the inverse Legendre map as:\begin{equation} \label{2.7bis} \hat\lambda^{-1} : \left\{
\begin{array}{l}
x^i = x^i \\
\\
a^\mu_i = a^\mu_i \\
\\
F^\mu_{ij} = \de H /de {\Pi_\mu^{ij}}
\end{array}
\right.
\end{equation}

\noindent Taking the Legendre map into account, as well as its inverse
\eqref{2.7bis}, it is easy to see that the image $\lambda\/(\je)$ defined by
eq.~\eqref{2.10bis} yields a Hamiltonian section $h\/$, represented by a function
$\H\/(x^i,a^\mu_i,\Pi_\mu^{ij}) = H(x^i,a^\mu_i,\Pi_\mu^{ij}) + \frac{1}{2}a^\nu_k a^\rho_r
C^\mu_{\rho\nu} \Pi_\mu^{kr}$.

\noindent Now, the presence of the Hamiltonian section allows to perform the pull-back of
the Liouville form on $\he$ to the phase space $\pe$. The result is a Hamiltonian
dependent $m-$form
\begin{equation}\label{2.7}
\Theta_h := h^*\/(\Theta) = -H(x^i,a^\mu_i,\Pi_\mu^{ij}) \; ds - \Pi_\mu^{ij}
\;\left( d\/a^\mu_{\,i}\wedge d\/s_j + \frac{1}{2} a^\nu_i a^\rho_j
C^\mu_{\rho\nu} ds \right)
\end{equation}

\noindent The variational principle constructed on the phase space $\pe$ with the
$m$-form $\Theta_h$ yields the Hamilton equations for the problem. In fact, the solution
of the variational problem for the functional
\[
A_h\/(\gamma) = \int_D \gamma^*\/(\Theta_h) \qquad \forall \; \rm{section} \; \gamma :
D\subset M \to \H\/(E)
\]
is made of its Euler--Lagrange equations
\begin{equation}\label{2.12} 
\gamma^*\/(X\interior d\Theta_h) = 0 \quad \forall\; X \in V\/(\pe,M)
\end{equation}
where $V\/(\pe,M)$ denotes the bundle of vectors over $\pe$ that are vertical with
respect to the fibration over $M$.

\noindent A straightforward calculation shows that eq.~\eqref{2.12} splits into
the following set of equations:
\begin{subequations}\label{2.13}
\begin{equation}
-\de H /de {\Pi_\mu^{ij}} - \de a^\mu_i /de {x^j} + \de a^\mu_j /de {x^i} -
a^\nu_i a^\rho_j C^\mu_{\rho\nu} = 0
\end{equation}
\begin{equation}
-\de H /de {a^\mu_{i}} - \de \Pi_\mu^{ji} /de {x^j} + \Pi_\lambda^{ji} a^\gamma_j
C^\lambda_{\gamma\mu} = 0
\end{equation}
\end{subequations}
usually referred to as {\it Hamilton--De Donder equations}.

The inverse Legendre map \eqref{2.7bis} shows that eq.~(\ref{2.13}a) is the
holonomy requirement for the solution, namely:
\[
F^\mu_{ij} = + \de a^\mu_j /de {x^i} - \de a^\mu_i /de {x^j} - a^\nu_i a^\rho_j
C^\mu_{\rho\nu}
\]

\noindent On the other hand, eq.~(\ref{2.13}b) can be written in terms of the
covariant derivative $D_j$ induced by the connection, giving rise to the usual
evolution equations for the Yang--Mills fields:
\[
D_j \Pi_\mu^{ji} = -\de H /de {a^\mu_{i}}
\]

%% file: par_3.tex
\noindent In the previous section a Hamiltonian approach to Yang--Mills field theories
has been developed, adapting the already known Hamilton--De Donder formalism developed
within the framework of calculus of variations to the new geometrical setting.

\noindent Nevertheless, there exists another well--known Hamiltonian approach to field
theory, represented by the so--called {\it multimomentum Hamiltonian formalism}, where
Hamiltonian connections play the same role as Hamiltonian vector fields in symplectic
geometry.

\noindent The argument has been widely studied in the literature (compare with \cite{Sardanashvily,Mangiarotti}), both
on the first jet--bundle and on the Legendre bundle (the phase space), in the Lagrangian and Hamiltonian
framework. In this section we will show that a multimomentum formulation of the above
theory can be built, starting from the Poincar\`e--Cartan forms \eqref{1.11} and
\eqref{2.7}.

\noindent We will start extending some definitions and some results about the Legendre
bundle of a generic field theory to our space. All the argument will be presented without
proofs; the reader is referred to \cite{Sardanashvily,Mangiarotti} for comments and further developments.

\noindent First of all, the canonical monomorphism is introduced as:
\[
    \Theta : \pe \hookrightarrow \Lambda^{m+1}\/T^*\/(E)\otimes_M T\/(M)
\]
\begin{equation}\label{3.1}
    \Theta := -\Pi^{ji}_\mu\,da^\mu_i \wedge ds \otimes \de /de{x^j}
\end{equation}

\noindent The following definitions are strictly associated with monomorphism
\eqref{3.1}.

\begin{Definition}
The pull--back valued horizontal form $\Theta$, locally described by eq.~\eqref{3.1} is
called multimomentum Liouville form on the phase space $\pe$.
\end{Definition}

\begin{Definition}
The pull--back valued form, defined as
\begin{equation}\label{3.2}
    \Omega :=  d\/\Pi^{ji}_\mu \wedge da^\mu_i \wedge ds \otimes \de /de{x^j}
\end{equation}
will be called the multisymplectic form on $\pe$.
\end{Definition}

\noindent The relation between the forms \eqref{3.1} and \eqref{3.2} is described by the
following
\begin{Proposition}
Given a generic 1--form $\sigma\in \Lambda^1\/(M)$, the forms \eqref{3.1} and \eqref{3.2}
are such that
\begin{equation}\label{3.3}
    \Omega \interior \sigma = -d\/(\Theta \interior \sigma)
\end{equation}
\end{Proposition}

\noindent Let us consider a connection $\gamma$ of the bundle $\pe\to M$, locally
described by the tangent--valued horizontal 1--form
\begin{equation}\label{3.4}
    \gamma = \left( \de /de{x^k} + \Gamma_{kh}^\mu\,\de /de{a^\mu_h} + \frac{1}{2}
    \Gamma_{k\mu}^{st}\,\de /de{\Pi_\mu^{st}} \right) \otimes d x^k
\end{equation}
where $\Gamma_{k\mu}^{st} = - \Gamma_{k\mu}^{ts}$.

\noindent Then, the following definition can be given:
\begin{Definition}
A connection $\gamma$ of the bundle $\pe\to M$, described by eq.~\eqref{3.4}, is called a
Hamiltonian connection if the $(m+1)$-form $\gamma\interior\Omega$ is closed.
\end{Definition}

\noindent A straightforward calculation shows that a connection $\gamma$ is Hamiltonian
if and only if it satisfies the following conditions:
\begin{equation}\label{3.5}
    \left\{
\begin{array}{l}
  \de \Gamma^{ji}_{j\sigma} /de {a^\lambda_p} =    \de \Gamma^{jp}_{j\lambda} /de {a^\sigma_i} = 0 \\
\\
  \de \Gamma^{ji}_{j\sigma} /de {\Pi_\lambda^{pq}} + \de \Gamma^{\lambda}_{pq} /de {a^\sigma_i} - \de \Gamma^{\lambda}_{qp} /de {a^\sigma_i} = 0 \\
\\
  \de \Gamma_{ji}^{\sigma} /de {\Pi_{\lambda}^{pq}} - \de \Gamma_{pq}^{\lambda} /de {\Pi_\sigma^{ji}}
  - \de \Gamma_{ij}^{\sigma} /de {\Pi_\lambda^{pq}} - \de \Gamma_{qp}^{\lambda} /de
  {\Pi_\sigma^{ji}}= 0
\end{array}
    \right.
\end{equation}

\begin{Definition}
An m--form $\eta\in\Lambda^1\/(\pe)$ is called a multimomentum Hamiltonian form if for
every open set $U \subset \pe$ there exists a Hamiltonian connection on $U$ satisfying
the equation
\begin{equation}\label{3.6}
    \gamma\interior \Omega = d \eta
\end{equation}
\end{Definition}

\noindent Now, we will show that the Poincar\`e--Cartan form \eqref{2.7} is a
multimomentum Hamiltonian form. In other words, we will show the existence of Hamiltonian
connections $\gamma$ satisfying the equation
\begin{equation}\label{3.7}
    \gamma\interior\Omega = d\Theta_h
\end{equation}

\noindent Moreover such connections will be shown to automatically satisfy the
Hamilton--De Donder equations \eqref{2.13}.

\noindent As a matter of fact, given a connection $\gamma$ in the form \eqref{3.4}, we
have that
\begin{equation}\label{3.9}
    \gamma \interior \Omega =\Gamma^{ji}_{j\sigma}\,d a^\sigma_i \wedge ds - \Gamma_{ji}^{\sigma}\,d\Pi_\sigma^{ji} \wedge
    ds + d\Pi^{ji}_{\sigma} \wedge d a^\sigma_{i} \wedge ds_j
\end{equation}

\noindent Nevertheless, one also as
\begin{equation}\label{3.10}
\begin{split}
    d \Theta_h = - \de H /de{a^\sigma_i}\,da^\sigma_i \wedge ds - \frac{1}{2}\de H
    /de{\Pi_\sigma^{ji}}\,d\Pi_\sigma^{ji} \wedge ds + d\Pi^{ji}_{\sigma} \wedge da^\sigma_{i}
     \wedge ds_j + \\
     +\frac{1}{2}a^\nu_{i}a^\rho_{j}C^\sigma_{\rho\nu}\,d\Pi^{ji}_\sigma \wedge ds +
     \Pi^{ji}_\mu C^\mu_{\rho\sigma} a^\rho_j\,da^\sigma_i \wedge ds
\end{split}
\end{equation}

\noindent A direct comparison of eqs.~\eqref{3.9} and \eqref{3.10} gives the algebraic
expressions satisfied by the components of $\gamma$:
\begin{subequations}\label{3.11}
\begin{equation}
\Gamma^{ji}_{j\sigma} +\de H /de{a^\sigma_i} - \Pi^{ji}_\mu C^\mu_{\rho\sigma} a^\rho_j =
0
\end{equation}
\begin{equation}
\Gamma_{ij}^\sigma - \Gamma_{ji}^\sigma + \de H /de{\Pi_\sigma^{ji}} - a^\nu_i a^\rho_j
C^\sigma_{\rho\nu} = 0
\end{equation}
\end{subequations}

\noindent Another direct comparison immediately shows that every integral section of a
connection $\gamma$ satisfying eqs.~\eqref{3.11} automatically verify the Hamilton--De
Donder eqs.~\eqref{2.13}.

\noindent The Lagrangian version of the above multimomentum Hamiltonian formulation is
obtained by means of Legendre transform. In fact the Lagrangian multisymplectic form on
$\je$ is defined through the Legendre map as:
\begin{equation}\label{3.12}
    \Omega_L := d\left( \de L /de{F^\mu_{ji}}\right) \wedge da^\mu_i \wedge ds  \otimes \de/de{x^j}
\end{equation}
The target connections of the fibration $\je\to M$ are of the form
\begin{equation}\label{3.13}
    \gamma = \left( \de /de{x^k} + \Gamma_{kh}^\mu\,\de /de{a^\mu_h} + \frac{1}{2}
    \Gamma^{\mu}_{kst}\,\de /de{F^\mu_{st}} \right) \otimes d x^k
\end{equation}
with $\Gamma^{\mu}_{kst} = - \Gamma^{\mu}_{kts}$, and satisfy the equation
\begin{equation}\label{3.14}
    \gamma\interior \Omega_L = d\Theta_L
\end{equation}
Because of the following relation
\[
\begin{split}
d\Theta_L = \de L /de{a^\mu_i}\,da^\mu_i \wedge ds - \frac{1}{2} F^\mu_{ij}\,d
\left( \de L /de{F^\mu_{ij}}\right) \wedge ds - d\left( \de L /de{F^\mu_{ij}}\right)
\wedge d a^\mu_{i} \wedge ds_j + \\
     - \frac{1}{2} a^\nu_i a^\rho_j C^\mu_{\rho\nu}\,d\left( \de L /de{F^\mu_{ij}}\right) \wedge ds
     - \de L /de{F^\mu_{ij}}C^\mu_{\rho\sigma} a^\rho_j\,da^\sigma_i \wedge ds
\end{split}
\]
it is easily seen that every $\gamma$ solution of eq.~\eqref{3.14} satisfies
the following conditions:
\begin{subequations}\label{3.15}
\begin{equation}
\begin{split}
\left( \frac{\partial^2 L }{\partial x^j \partial F^\sigma_{ji}} + \Gamma^\mu_{jh}
\frac{\partial^2 L }{\partial a^\mu_k \partial F^\sigma_{ji}} + \frac{1}{2}
\Gamma^\mu_{jst} \frac{\partial^2 L }{\partial F^\mu_{st} \partial F^\sigma_{ji}} \right)
 - \de L /de{F^\mu_{ji}} a^\gamma_j C^\mu_{\gamma\sigma} - \de L /de{a^\sigma_i} = 0
\end{split}
\end{equation}

\begin{equation}
F^\mu_{ij} +\Gamma^\mu_{ji} - \Gamma^\mu_{ij} + a^\lambda_i a^\gamma_j C^\mu_{\gamma\lambda} = 0
\end{equation}
\end{subequations}

\noindent Once again, it is easy to verify that the integral sections of such a connection
$\gamma$ automatically satisfy Euler--Lagrange equations \eqref{1.13}.

%% file: par_4.tex
\noindent In this section a particular Gauge theory is considered: the base manifold $M$
will be taken to be $3$-dimensional and the gauge groups can be equivalently chosen
between $G=SO\/(3)$ and $G=SO\/(2,1)$.

\noindent Under these hypotheses, both space--time and algebra indexes run from $1$ to
$3$. Besides, given a basis $\{\underbar{e}_\mu\}$ for the Lie algebra $\frak{g}$ of $G$,
we denote by $K_{\mu\nu}$ the coefficients of an Ad-invariant metric over $\frak{g}$
such that the structure coefficients $C^\mu_{\;\;\lambda\sigma}$ are expressed in the
form
\begin{equation}\label{5.1}
    C^\mu_{\;\;\lambda\sigma} = \frac{1}{2}\sqrt{K} K^{\mu\nu}
    \epsilon_{\nu\lambda\sigma}
\end{equation}
where $\sqrt{K} = \sqrt{|\det K_{\mu\nu}|}$, $K^{\mu\nu}K_{\nu\sigma} =
\delta^\mu_\sigma$ and $\epsilon_{\nu\lambda\sigma}$ are the $3$-dimensional Levi--Civita
permutation symbols.

\noindent The use of a {\it dual formulation} \cite{Ferraris} allows to express such a
$(3+3)$ gauge theory in terms of a gravity--like theory in purely metric formulation, as
proved in \cite{Raiteri1,Raiteri2}.

\noindent Now, making an explicit use of the Poincar\'e--Cartan approach of section 3, we
will show that the Hamiltonian version of a $(3+3)$ gauge theory has the same shape as an
Einstein--Cartan theory. Borrowing from \cite{Raiteri1,Raiteri2}, the central idea
consists in performing a local coordinate transformation in the phase space $\pe$,
locally described by the following relations:
\begin{equation}\label{5.2}
    \left\{
\begin{array}{l}
e^\nu_p = \frac{1}{2} K^{\mu\nu}\Pi_\mu^{ij} \epsilon_{pij} \\
\\
\omega_{i\beta\alpha} = \frac{1}{2} \sqrt{K} \epsilon_{\mu\alpha\beta} a^\mu_i
\end{array}
\right.
\end{equation}
where $\epsilon$ denotes the usual $3$-dimensional Levi--Civita permutation symbol. The inverse
transformation of \eqref{5.2} is given by
\begin{equation}\label{5.3}
\left\{
\begin{array}{l}
    \Pi_\mu^{ij} = K_{\mu\nu} e^\nu_p \epsilon^{pij} \\
\\
    a^\mu_i = \frac{1}{\sqrt{K}} \epsilon^{\mu\sigma\lambda}\omega_{i\lambda\sigma}
\end{array}
\right.
\end{equation}

\noindent It will soon be clear that the coordinates $e^\mu_i$ play the role of the triad
coordinates, while the coordinates $\omega_{i\beta\alpha}$ represent the coefficients of
the spin--connection.

\noindent It is now easy to see that the Poincar\'e--Cartan 1--form \eqref{2.7} in the
new coordinates has the form
\begin{equation}\label{5.4}
    \Theta_h = - H \/ ds - K_{\mu\nu} e^\nu_p\epsilon^{pij} \frac{1}{\sqrt{K}}
    \left(\epsilon^{\mu\sigma\lambda}\/d\omega_{i\lambda\sigma}\wedge ds_j +
    \frac{1}{2}\epsilon^{\sigma\alpha\beta}\omega^{\;\;\mu}_{j\;\;\sigma}\/\omega_{i\beta\alpha}
    ds \right)
\end{equation}
where $\omega^{\;\;\mu}_{j\;\;\;\sigma} := \omega_{j\nu\sigma}K^{\mu\nu}$.

\begin{Proposition}
The following identities hold identically:
\begin{equation}\label{5.5}
    - \frac{1}{2} \epsilon^{\rho\alpha\beta}\omega_{j\nu\rho}\omega_{i\beta\alpha} =
    K_{\mu\nu}\epsilon^{\mu\sigma\lambda}\omega_{i\lambda\eta}\omega_{j\;\;\;\sigma}^{\;\;\eta}
\end{equation}
\end{Proposition}
\demo A direct calculation shows that the left hand side is such that
\[
    - \frac{1}{2} \epsilon^{\rho\alpha\beta}\omega_{j\nu\rho}\omega_{i\beta\alpha} =
     \omega_{j\nu 1}\omega_{i23} - \omega_{j\nu 2}\omega_{i13} +
\omega_{j\nu 3}\omega_{i12}
\]
while the right hand side becomes
\[
\begin{split}
K_{\mu\nu} 
 \epsilon^{\mu\sigma\lambda}\omega_{i\lambda\eta}\omega_{j\;\;\;\sigma}^{\;\;\eta} =
K_{\nu 1} \omega_{i3\eta}\omega_{j\;\;\;2}^{\;\;\eta} - K_{\nu 2}
\omega_{i3\eta}\omega_{j\;\;\;1}^{\;\;\eta} + K_{\nu 3}
\omega_{i2\eta}\omega_{j\;\;\;1}^{\;\;\eta} + \\
- K_{\nu 1} \omega_{i2\eta}\omega_{j\;\;\;3}^{\;\;\eta} + K_{\nu 2}
\omega_{i1\eta}\omega_{j\;\;\;3}^{\;\;\eta} - K_{\nu 3}
\omega_{i1\eta}\omega_{j\;\;\;2}^{\;\;\eta} = \\
K_{\nu 1} \omega_{i31}\omega_{j\;\;\;2}^{\;\;1} + K_{\nu 1}
\omega_{i32}\omega_{j\;\;\;2}^{\;\;2} - K_{\nu 2}
\omega_{i31}\omega_{j\;\;\;1}^{\;\;1} - K_{\nu 2}
\omega_{i32}\omega_{j\;\;\;1}^{\;\;2} + \\
K_{\nu 3} \omega_{i21}\omega_{j\;\;\;1}^{\;\;1} + K_{\nu 3}
\omega_{i23}\omega_{j\;\;\;1}^{\;\;3} - K_{\nu 1}
\omega_{i21}\omega_{j\;\;\;3}^{\;\;1} - K_{\nu 1}
\omega_{i23}\omega_{j\;\;\;3}^{\;\;3} + \\
 K_{\nu 2} \omega_{i12}\omega_{j\;\;\;3}^{\;\;2} + K_{\nu 2} 
\omega_{i13}\omega_{j\;\;\;3}^{\;\;3} - K_{\nu 3} 
\omega_{i12}\omega_{j\;\;\;2}^{\;\;2} - K_{\nu 3}
\omega_{i13}\omega_{j\;\;\;2}^{\;\;3}
\end{split}
\]

\noindent Now we notice that:
\[
\omega_{j\nu 1} \omega_{i23} = 
K_{\nu1}\omega_{i23}\omega_{j\;\;\;1}^{\;\;1} + K_{\nu2}\omega_{i23}\omega_{j\;\;\;1}^{\;\;2}
+ K_{\nu3}\omega_{i23}\omega_{j\;\;\;1}^{\;\;3}
\]
\[
\omega_{j\nu 2}\omega_{i13} = K_{\nu
1}\omega_{i13}\omega_{j\;\;\;3}^{\;\;1} + K_{\nu 2} \omega_{i13}\omega_{j\;\;\;3}^{\;\;2} +
K_{\nu 3} \omega_{i13}\omega_{j\;\;\;3}^{\;\;3}
\]
\[
\omega_{j\nu 3}\omega_{i12} = K_{\nu
1}\omega_{i12}\omega_{j\;\;\;2}^{\;\;1} + K_{\nu 2}\omega_{i12}\omega_{j\;\;\;2}^{\;\;2} +
K_{\nu 3}\omega_{i12}\omega_{j\;\;\;2}^{\;\;3} 
\]
whence:
\[
\begin{split}
K_{\mu\nu}
 \epsilon^{\mu\sigma\lambda}\omega_{i\lambda\eta}\omega_{j\;\;\;\sigma}^{\;\;\eta} &=
- \frac{1}{2}
\epsilon^{\rho\alpha\beta}\omega_{j\nu\rho}\omega_{i\beta\alpha} + \\
&- K_{\nu1}\omega_{i23}\omega_{j\;\;\;\mu}^{\;\;\mu} -
K_{\nu2}\omega_{i31}\omega_{j\;\;\;\mu}^{\;\;\mu} -
K_{\nu3}\omega_{i12}\omega_{j\;\;\;\mu}^{\;\;\mu}
\end{split}
\]
The conclusion follows from the trace properties of the coefficients
$\omega_{j\;\;\;\mu}^{\;\;\mu} = 0$.
\enddemo

\noindent Taking the identity \eqref{5.5} into account, we can write the differential of
$\Theta_h$ in the form:
\begin{equation}\label{5.6}
\begin{split}
    d\Theta_h =& - \de H /de{e^\lambda_i}\,de^\lambda_i \wedge ds - \frac{1}{2} \de H
    /de{\omega_{i\lambda\sigma}}\,d\omega_{i\lambda\sigma} \wedge ds +
    \frac{1}{\sqrt{K}}
    \epsilon^{pij}\epsilon^{\rho\alpha\beta}\omega_{j\rho\nu}e^\nu_p\,
    d\omega_{i\beta\alpha}\wedge ds \\
    &- K_{\mu\nu} \epsilon^{pij}\,de^\nu_p \wedge \frac{1}{\sqrt{K}}
     \epsilon^{\mu\sigma\lambda}\,\left(d \omega_{i\lambda\sigma} \wedge ds_j -
    \omega_{i\lambda\eta}\omega_{j\;\;\;\sigma}^{\;\;\eta}\,ds \right)
    \end{split}
\end{equation}

\noindent Now, let $X = X^\nu_p\,\de /de{e^\nu_p} + \frac{1}{2} X_{i\lambda\sigma}\,\de
/de{ \omega_{i\lambda\sigma}}$ be a vertical vector field, with respect to the fibration
$\pe\to M$, on the phase space $\pe$. We calculate the inner product
\begin{equation}\label{5.7}
    \begin{split}
X\interior d\Theta_h =& \left( - \de H /de{e^\nu_p}\,ds - \epsilon^{pij}
\epsilon^{\mu\sigma\lambda} \frac{K_{\mu\nu}}{\sqrt{K}}\,d\omega_{i\lambda\sigma} \wedge
ds_j + \epsilon^{pij} \epsilon^{\mu\sigma\lambda} \frac{ K_{\mu\nu}}{\sqrt{K}}
\omega_{i\lambda\eta}\omega_{j\;\;\;\sigma}^{\;\;\eta}\,ds \right) X^\nu_p \\
+& \left( - \frac{1}{2} \de H /de{\omega_{i\lambda\sigma}}\,ds + \epsilon^{pij}
\epsilon^{\mu\sigma\lambda} \frac{ K_{\mu\nu}}{\sqrt{K}}\,de^\nu_p \wedge ds_j +
\frac{1}{\sqrt{K}}
    \epsilon^{pij}\epsilon^{\rho\sigma\lambda}\omega_{j\rho\nu} e^\nu_p\,ds \right) X_{i\lambda\sigma}
    \end{split}
\end{equation}

\noindent The imposition on the Hamilton--De Donder conditions $\gamma^*\/(X\interior
d\Theta_h) = 0\; \forall\; X$ yields the final equations
\begin{subequations}\label{5.8}
    \begin{equation}
    - \de H /de{e^\lambda_i} - \epsilon^{pij} \epsilon^{\mu\sigma\lambda} \frac{K_{\mu\nu}}{\sqrt{K}}
    \left(\de \omega_{i\lambda\sigma} /de{x^j} + \omega_{j\lambda\eta}\omega_{i\;\;\;\sigma}^{\;\;\eta}
    \right) = 0
    \end{equation}
    \begin{equation}
    - \de H /de{\omega_{i\lambda\sigma}} + \frac{2K_{\mu\nu}}{\sqrt{K}}
    \epsilon^{pij}\epsilon^{\mu\sigma\lambda} \left( \de e^\nu_p /de{x^j} +
    \omega_{j\;\;\;\gamma}^{\;\;\nu} e^\gamma_p \right) = 0
    \end{equation}
\end{subequations}
representing the Hamilton--De Donder equations in the new coordinates.

\noindent As it was anticipated at the beginning of the section, eqs.~\eqref{5.8} have
the form of the $3$-dimensional Einstein--Cartan equations, where the coordinates
$e^\mu_i$ and $\omega_{i\mu\nu}$ respectively represent the triad components (whenever
$\det \| e^\mu_i\| \neq 0$) and the spin--connection coefficients.

\noindent In particular, let us consider a free Yang--Mills field, whose dynamical
properties are described by the usual Lagrangian density $L= - \frac{1}{4} F^\mu_{ip}
F^\nu_{jq} g^{ij} g^{pq} K_{\mu\nu} \sqrt{g}$, where $g_{ij}$ is a given metric over $M$ and $g:=|\det g_{ij}|\/$.
Under such circumstances, the Legendre transformation and the Hamiltonian are
respectively described by the following equation:
\[
\Pi_\mu^{ij} = \de L /de{F^\mu_{ij}} = -F^\nu_{pq} g^{ip} g^{jq} K_{\mu\nu} \sqrt{g} \quad,
\quad H = - \frac{1}{4}\frac{1}{\sqrt{g}} \Pi_\sigma^{pq} \Pi_\lambda^{st} g_{sp}g_{tq}
K^{\sigma\lambda}
\]

\noindent When the new coordinates \eqref{5.2} are introduced, the Hamiltonian takes the
form:
\[
H = - \frac{1}{2} G_{kh} g^{kh} \sqrt{g} \sigma\/(g) \qquad (G_{hk} := e^\mu_k e^\nu_h
K_{\mu\nu})
\]
with $\sigma\/(g)$ representing the sign of $\det \|g_{ij}\|$. Since
\[
\de H /de{\omega_{i\lambda\sigma}} = 0 \quad ; \quad \de H /de{e^\nu_p} = - e^\mu_k
K_{\mu\nu} g^{kp} \sqrt{g} \sigma\/(g)
\]
eqs.~\eqref{5.8} take the form
\begin{subequations}\label{5.9}
    \begin{equation}
    2 K_{\mu\nu}\epsilon^{pij}\epsilon^{\mu\sigma\lambda}\left( \de e^\nu_p /de{x^j} +
    \omega_{j\;\;\;\gamma}^{\;\;\nu} e^\gamma_p  \right) = 0
    \end{equation}
    \begin{equation}
    \frac{1}{2} e^\mu_k K_{\mu\nu} g^{kp} \sqrt{g} \sigma\/(g) -
    \epsilon^{pij}\epsilon_{\nu\lambda\sigma} R_{ij}^{\;\;\;\;\lambda\sigma}\sqrt{K}
    \sigma(K) = 0
    \end{equation}
\end{subequations}
where
\[
R_{ij\lambda\sigma} = \de \omega_{j\lambda\sigma} /de{x^i} - \de \omega_{i\lambda\sigma}
/de{x^j} + \omega_{i\lambda\eta} \omega_{j\;\;\;\sigma}^{\;\;\eta} - \omega_{j\lambda\eta}
\omega_{i\;\;\;\sigma}^{\;\;\eta} \quad , \quad R_{ij}^{\;\;\;\;\lambda\sigma} =
R_{ij\mu\nu} K^{\mu\lambda}K^{\nu\sigma}
\]
and $\sigma\/(K) = \rm{sign}\/(\det\|K_{\mu\nu}\|))$.

\noindent Under the hypothesis $\det \|e^\mu_i\| \neq 0$ eqs.~\eqref{5.9} have the same
form as Einstein equations in the triad--affine formulation. Because of eq.~(\ref{5.9}a),
the solution $\omega_{i\mu\nu}\/(x)$ is equal to the (spin--connection associated with) Levi--Civita connection induced by
the metric $G = K_{\mu\nu} e^\mu\/(x) \otimes e^\nu\/(x)$, which is a solution of
eq.~(\ref{5.9}b).

\noindent More in particular, eqs.~\eqref{5.9} actually describe a first--order purely 
frame--formulation
of a General Relativity like theory in three dimensions.

\noindent Infact, we notice that the transformation laws of the coordinates \eqref{5.2} are 
\begin{subequations}
\begin{equation}\label{5.10}
\bar{e}^\mu_j = e^\sigma_{i}Ad\/(\gamma^{-1})^\mu_{\sigma}\de x^i/de{\bar{x}^j}
\end{equation}
and
\begin{equation}\label{5.11}
\bar{\omega}_{i\mu\nu}=Ad\/(\gamma)_\mu^{\sigma}Ad\/(\gamma)_{\nu}^{\gamma}\de
x^j/de{\bar{x}^i}\omega_{j\sigma\gamma} + Ad\/(\gamma)_{\mu}^{\eta} \de
Ad\/(\gamma^{-1})^\sigma_{\eta}/de{x^h}\de x^h/de{\bar{x}^i}K_{\sigma\nu}
\end{equation}
\end{subequations}
Eqs.~\eqref{5.10} are the transition functions of a bundle $\pi: {\cal T}\to M\/$, associated with
$P\times_M L\/(M)\/$ ($L\/(M)\/$ being the frame bundle over $M\/$) through the left action
\begin{equation}\label{5.12}
\lambda : (G \times GL\/(3,\Re))\times GL\/(3,\Re)\to GL\/(3,\Re),\quad
\lambda\/(g,J;X):=Ad\/(g)\cdot X\cdot J^{-1}
\end{equation}
The (local) sections $e:M\to {\cal T}\/$ may be identified with (local) triads $e^\mu_i\/(x)\,dx^i\/$ on $M\/$; the latter are truly gauge natural objects \cite{FF}, sensitive to the changes of trivialization of the structure bundle $P\/$. Each triad $e^\mu\/$ induces a metric on $M\/$ expressed as $G:= K_{\mu\nu}e^\mu \otimes e^\nu\/$, which is invariant under transformations \eqref{5.10} by construction. 

\noindent
A new $\cal J\/$-bundle $\hat{\pi}: {\cal J}\/({\cal T})\to M\/$ can also be constructed by quotienting the first--jet bundle $j_1\/({\cal T})\/$ of $\pi: {\cal T}\to M\/$
with respect to an equivalence relation
analogous to \eqref{1.2tris}. The bundle ${\cal J}\/({\cal T})\/$ is naturally referred to local coordinates $x^i,e^\mu_i,E^\mu_{ij}:=\frac{1}{2}\left(e^\mu_{ij} - e^\mu_{ji}\right)\/$ $(i<j)\/$.

\noindent Now the idea is to choose the components of the {\it spin--connections} generated by the triads
themselves as fiber coordinates on the bundle ${\cal J}\/({\cal T})\/$. 

\noindent
Within this framework, let $z=(x^i,e^\mu_i,E^\mu_{ij})\/$
be an element of ${\cal J}\/({\cal T})$, $x=\hat\pi\/(z)$ its projection over $M$, $e^\mu$ a
representative triad belonging to the equivalence class $z$ and $G=K_{\mu\nu} e^\mu\otimes e^\nu$ the metric on $M$ 
induced by the triad $e^\mu$; we also denote by $\Gamma_{ih}^k$ the Levi--Civita connection induced by the metric $G\/$ and 
by $\omega_{i\;\;\;\nu}^{\;\;\mu}\/$ the spin connection associated with $\Gamma_{ih}^k$ through the triad $e^\mu\/$ itself. 

\noindent The relation between the coefficients $\Gamma_{ih}^k$ and $\omega_{i\;\;\;\nu}^{\;\;\mu}$,
evaluated in the point $x=\hat \pi\/(z)\in M$, is expressed by the equation
\begin{equation}\label{5.12bis}
\omega^{\;\;\mu}_{i\;\;\;\nu}\/(x) = e^\mu_k\/(x)\left( \Gamma^k_{ij}e^j_\nu\/(x) +
\de{e^k_\nu\/(x)}/de{x^i} \right)
\end{equation}
If the coefficients
$\Gamma_{ih}^k$ are written in terms of the triad $e^\mu$ and its derivatives, one gets
the well--known expression
\begin{equation}\label{5.13}
\omega^{\;\;\mu}_{i\;\;\;\nu}\/(x):= e^\mu_p\/(x) \left( \Sigma^p_{\;\;ji}\/(x) -
\Sigma_{j\;\;i}^{\;\;p}\/(x) + \Sigma_{ij}^{\;\;\;p}\/(x) \right) e^j_{\nu}\/(x)
\end{equation}
where
\begin{equation}\label{5.14}
\Sigma^p_{\;\;ji}\/(x):= e^p_\lambda\/(x) E^\lambda_{ij}\/(x) =
e^p_\lambda\/(x)\frac{1}{2}\left( \de{e^\lambda_i\/(x)}/de{x^j} -
\de{e^\lambda_j\/(x)}/de{x^i} \right)
\end{equation}
the Latin indexes being lowered and raised by means of the metric $G$. Equations \eqref{5.13} and \eqref{5.14} show that the values of the coefficients
of the spin--connection $\omega_{i\;\;\;\nu}^{\;\;\mu}$, evaluated in $x=\hat \pi\/(z)$,
are independent of the choice of the representative $e^\mu$ in the equivalence class
$z\in {\cal J}\/({\cal T})$.

\noindent
Moreover, the torsion--free condition for the connection
$\omega_{i\;\;\;\nu}^{\;\;\mu}$ gives a sort of inverse relation of eq.~\eqref{5.13} in the form
\begin{equation}\label{5.15}
2E^\mu_{ij}\/(x) = \omega^{\;\;\mu}_{i\;\;\;\nu}\/(x)e^\nu_j\/(x) -
\omega^{\;\;\mu}_{j\;\;\;\nu}\/(x)e^\nu_i\/(x)
\end{equation}

\noindent
Because of the metric compatibility condition $\omega_{i\mu\nu}:=
\omega_{i\;\;\;\nu}^{\;\;\sigma}K_{\sigma\mu} = - \omega_{i\nu\mu}$, there exists
a one-to-one correspondence between the values of the antisymmetric part of the
derivatives $E^\mu_{ij}\/(x) = \frac{1}{2} \left( \de e^\mu_{i}\/(x) /de{x^j} - \de e^\mu_{j}\/(x)
/de{x^i}\right)$ and the coefficients of the spin--connection $\omega_{i\mu\nu}\/(x)$ in
the point $x=\hat\pi\/(z)$.

\noindent
The above considerations allow us to take the quantities
$\omega_{i\mu\nu}\/$ as fiber coordinates of the bundle ${\cal J}\/({\cal T})\/$, looking at the
relations \eqref{5.13} and \eqref{5.15} as coordinate changes in ${\cal J}\/({\cal T})\/$.

\noindent
Finally, it is a straightforward matter to verify that the transformation laws of the spin connection coefficients
$\omega_{i\mu\nu}\/$ coincide with eqs.~\eqref{5.11}, as well as that the $3$-form \eqref{5.3} is invariant under coordinate transformations
\eqref{5.10}, \eqref{5.11}.